\documentclass[a4paper,UKenglish,cleveref, autoref, thm-restate]{lipics-v2021}

\hideLIPIcs  

\usepackage{wrapfig}
\usepackage{subcaption}
\usepackage{tikz}
\usetikzlibrary{shapes}
\usepackage{pgfplots}
\pgfplotsset{width=10cm,compat=1.9}
\usetikzlibrary{automata, arrows.meta, positioning, matrix,arrows,arrows.meta,fit}
\usetikzlibrary{decorations.pathmorphing}
\usepackage{caption}
\usepackage{xcolor}
\usepackage{float}
\usepackage{todonotes}
\usepackage[ruled,noend]{algorithm2e}
\usepackage{tabularx}

\lstdefinestyle{compactjson}{
  language=json,
  basicstyle=\ttfamily\scriptsize,
  columns=fullflexible,
  keepspaces=true,
  breaklines=true,
  aboveskip=4pt,
  belowskip=4pt
}

\usepackage{rotating}
\usepackage{mathbbol}
\usepackage{lscape}

\usepackage{amssymb}
\usepackage{indentfirst}
\usepackage{stmaryrd}
\usepackage{array,multirow}

\usepackage{svg}
   \nolinenumbers
   
\definecolor{darkgreen}{rgb}{0.0, 0.56, 0.0}

\title{\textsc{PaSTTeL}: Parallel analysiS framework for Termination and non-Termination of Lasso programs}

\author{Anissa Kheireddine}{Sorbonne University, LIP6, France\and Dowsers, France \and \url{https://akheireddine.github.io/} }{anissa.kheireddine@lip6.fr}{[orcid]}{}{}

\author{Souheib Baarir}{Sorbonne University, LIP6, France,\and Paris-Nanterre University, France  \and  \url{https://perso.lip6.fr/Souheib.Baarir/}}{souheib.baarir@lip6.fr}{[orcid]}{}{}

\author{Hugo De Sa Pereira Pinto}{Dowsers, France}{Hugo@dowsers.finance}{}{}

\authorrunning{A. Kheireddine et al.} 




\category{} 

\relatedversion{} 

\EventEditors{John Q. Open and Joan R. Access}
\EventNoEds{2}
\EventLongTitle{42nd Conference on Very Important Topics (CVIT 2016)}
\EventShortTitle{CVIT 2016}
\EventAcronym{CVIT}
\EventYear{2016}
\EventDate{December 24--27, 2016}
\EventLocation{Little Whinging, United Kingdom}
\EventLogo{}
\SeriesVolume{42}
\ArticleNo{23}

\begin{document}
\tikzset{
    ->, 
    node distance=4cm, 
    transform shape,
    every state/.style={draw=none,fill=cyan!10,scale=1.2}, 
    initial text=$ $, 
}

\maketitle

\begin{abstract}

Proving termination or non-termination of \emph{lasso programs} is a challenging problem in program verification. To unify state-of-the-art approaches under a common execution framework, we present \textsc{PaSTTeL}, a modular and generic parallel portfolio framework for termination and non-termination analysis of lasso programs. \textsc{PaSTTeL} is designed to: (1) facilitate the integration of new analysis algorithms into the portfolio, (2) execute registered strategies concurrently, and (3) act as a self-contained library component that can be seamlessly embedded into any external project requiring (non-)termination analysis. Initial experiments demonstrate that an instantiation of \textsc{PaSTTeL} performs competitively against state-of-the-art tools.

\end{abstract}

\section{Introduction}

Proving the termination of a program is one of the most fundamental correctness properties in computer science, and is known to be undecidable in the general case. A practical way to approach this problem is to search for explicit termination or non-termination arguments. The existence of a certain termination argument — such as a linear ranking function — is decidable~\cite{BradleyMS05,ranking-template2015,PodelskiR04} and implies termination. However, if no linear ranking function can be found, we cannot conclude non-termination. Conversely, the existence of a non-termination argument, e.g., a linear recurrence set~\cite{recurrentsets2008} or a geometric non-termination argument~\cite{LeikeH16}, is decidable and implies non-termination, but a failure to find one cannot conclude termination.
These analyses can be applied to \emph{lasso programs}: execution traces consisting of a single while loop preceded by straight-line code, whose control flow graph takes the characteristic shape of a lasso. In typical use cases, a lasso program is extracted from a control flow graph to represent a candidate infinite execution path, as encountered in CEGAR-based termination synthesis~\cite{BryonHN15,BryonEM11,abstractRefinement05,DietschHLP15}.

In this work, we introduce \textsc{PaSTTeL}\footnote{\url{https://github.com/akheireddine/PaSTTeL.git} \label{fn:pasttel}}, a parallel analysis framework for the termination and non-termination of linear lasso programs (\textit{llp}). \textsc{PaSTTeL} unifies state-of-the-art techniques within a concurrent portfolio that combines both termination and non-termination approaches. Our main contributions are as follows:
\begin{itemize}
    \item[--] We propose a modular and generic framework for implementing termination and non-termination strategies, designed to facilitate their integration into a unified portfolio.
    
   \item[--] We implement several state-of-the-art constraint-based techniques, including ranking function synthesis~\cite{ranking-template2015} and geometric non-termination arguments~\cite{LeikeH16}.
  
    \item[--] We evaluate an instance of our framework on the \textsc{SV-COMP 2025}\footnote{\url{https://sv-comp.sosy-lab.org/2025/}\label{fn:svcomp25}} benchmark suite, comparing its performance with the state-of-the-art \textsc{Ultimate LassoRanker (ULR)}\footnote{\url{https://www.ultimate-pa.org/lasso_ranker/}\label{fn:lassoranker}}. The results show that the modularity and genericity of \textsc{PaSTTeL} do not compromise performance, while its C++-based implementation yields a significant speedup in execution time.
\end{itemize}
\noindent Beyond its standalone use, \textsc{PaSTTeL} is designed as a self-contained library that any external project can integrate to leverage (non-)termination analysis capabilities without knowledge of its internal design.

\section{Architecture of the Framework}
\label{sec:framework}

\begin{figure}[!htp]
    \centering
    \includegraphics[scale=0.77]{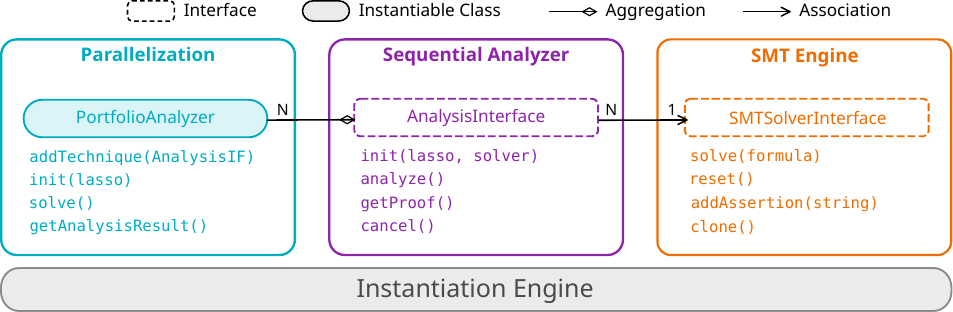}
    \caption{The architecture of \textsc{PaSTTeL}}
    \label{fig:framework}
\end{figure}

\textsc{PaSTTeL} is a generic, modular, and efficient framework developed in \textsc{C++17}, enabling straightforward implementation and parallel execution of termination and non-termination synthesis. The architecture of \textsc{PaSTTeL} is structured around three core components: a sequential analyzer (\texttt{AnalysisInterface}); an SMT engine (\texttt{SMTSolverInterface}); and a parallelization layer (\texttt{PortfolioAnalyzer}), as illustrated in Figure~\ref{fig:framework}). Any instantiation of the framework must provide concrete implementations of these three components. 
\begin{description}

    \item \textbf{SMT engine:} \texttt{SMTSolverInterface}. This interface allows different SMT solvers to be plugged into the framework interchangeably. \textsc{PaSTTeL} currently supports \textsc{Z3} and \textsc{CVC5}. The solver instance is passed to each sequential analyzer and used by the constraint-based analysis techniques.

    \item \textbf{Sequential Analyzer:} \texttt{AnalysisInterface}. This adapter defines the basic functions expected from a sequential termination or non-termination analyzer. The main methods of this interface are:
        \begin{itemize}
            \item[--] \texttt{AnalysisResult analyze()}: launches the analysis using the analyzer's own SMT solver instance. Returns \texttt{TERMINATING}, \texttt{NON\_TERMINATING}, or \texttt{UNKNOWN}.
            \item[--] \texttt{ProofCertificate getProof()}: returns a proof certificate for the (non-)termination result in plain text format.
        \end{itemize}
    
    \item \textbf{Parallelization:} \texttt{PortfolioAnalyzer.} This component manages the threads deployed for the concurrent execution of termination and non-termination strategies on the same \textit{llp} input. Its main methods are:
    \begin{itemize}
        \item[--] \texttt{void addTechnique(AnalysisInterface* technique)}: registers an analysis technique into the portfolio.
        \item[--] \texttt{void solve()}: launches the concurrent search across all registered strategies.
        \item[--] \texttt{ProofCertificate join(int timelimit)}: blocks until one technique returns a conclusive result or the time limit is reached. If the result differs from \texttt{UNKNOWN}, the corresponding proof certificate is returned.
    \end{itemize}
\end{description}

\begin{algorithm}[tp]
\caption{\texttt{init-PaSTTeL} function}
\label{fig:algo}
\KwIn{\textit{lasso}: \textit{llp}, \textit{n}: number of threads, \textit{solverType}: \{\texttt{Z3}, \texttt{CVC5}\}}
\textit{solvers} $\leftarrow$ Create \textit{n} of \textit{solverType} from \texttt{SMTSolverInterface}\;
\textit{portfolio} $\leftarrow$ Create \texttt{PortfolioAnalyzer(\textit{lasso},\textit{n})}\;
\textit{portfolio}\texttt{.addTechnique(new TerminationSA(...), \textit{solvers[0]})}\;
\textit{portfolio}\texttt{.addTechnique(new NonTerminationSA(...),  \textit{solvers[1]})}\;
\textit{portfolio}\texttt{.solve()}\;
\textit{portfolio}\texttt{.join(}\textit{timelimit}\texttt{)}\;
\texttt{print(}\textit{portfolio}\texttt{.getAnalysisResult())}\;
\If{\textit{portfolio}\texttt{.getAnalysisResult()} $\neq$ \textsc{Unknown}}{
    \texttt{print(}\textit{portfolio}\texttt{.getProof())}\;
}
\end{algorithm}

\noindent\textbf{Engine Instantiation.} To create a particular instance of \textsc{PaSTTeL}, the user has to adapt the \texttt{init-PaSTTeL} function (Algorithm~\ref{fig:algo}), which serves as the entry point for instantiating and binding all components of the framework. Its parameters expose the key configuration points, allowing any external tool to embed \textsc{PaSTTeL} as a self-contained library with the desired configuration.

\section{\textsc{PaSTTeL} "à la \textsc{ULR}"}
\label{sec:implementation}


To validate the generic aspect of our framework, we selected the techniques underlying  \textsc{Ultimate LassoRanker} (\textsc{ULR}) --- one of the state-of-the-art tool for linear lasso-level analysis and overall winner of \textsc{SV-COMP 2025} --- as a proof of concept. We implemented \textsc{P-ULR}, an instantiation of \textsc{PaSTTeL} that replicates the core (non-)termination strategies of \textsc{ULR}. Each technique is realized as an independent analyzer registered into the portfolio and executed concurrently with the others. 
\\

\noindent\textbf{Input Format \& Linearization.} While \textsc{PaSTTeL} operates directly on \textit{llp}, \textsc{P-ULR} provides a JSON-based input format as a convenience layer, designed to be generic and tool-agnostic. Formulas are encoded following the SMT-LIB standard, with support for Single Static Assignment (SSA) form. 

Given a lasso program in the JSON-based SMT-LIB format, \textsc{P-ULR} first linearizes it, if necessary, then dispatches the analysis to the registered strategies. 
\\
\noindent\textbf{Termination.} \textsc{P-ULR} provides a constraint-based synthesis of termination arguments for \textit{llp} using linear ranking templates~\cite{ranking-template2015}: it implements \textit{affine} and \textit{nested} ranking templates\footnote{integrating additional templates such as \textit{lexicographic}, \textit{multiphase}, or \textit{piecewise} requires few lines of \textsc{C++} code}. 

\noindent\textbf{Non-Termination.} \textsc{P-ULR} implements a constraint-based synthesis of non-termination arguments for \textit{llp}: it implements \textit{fixpoint} checking and \textit{Geometric Non-Termination Arguments} (\textit{GNTA})~\cite{LeikeH16}.
\\

\noindent\textbf{Limitations.} The current version of \textsc{P-ULR} has several limitations. First, the linearization step does not yet cover all non-linear cases: in particular, array operations may require special treatment, including the generation of supporting invariants for array variables, which is not currently supported. Second, at the framework level, the lasso program structure currently provides better support for linear formulas; non-linear formulas are passed as-is to the registered strategies, which is valid for strategies that do not require linearization such as \textit{fixpoint}. Third, at the strategy level, \textsc{P-ULR} currently supports linear rational arithmetic (\textsc{QF\_LRA}) for termination synthesis and linear integer arithmetic (\textsc{QF\_LIA}) for \textit{GNTA} synthesis. Finally, simplification of synthesized ranking function coefficients and formula optimization have not yet been implemented. Extensions to \textsc{QF\_NRA} and \textsc{QF\_NIA}, as well as the improvements mentioned above, are left for future work.

\section{Numerical comparison}
\label{sec:exp}

\begin{table}[tp]
    \centering
    \begin{tabularx}{1\textwidth}{X|c|c||c|c}
         \textbf{Tool} & \textbf{\#Terminating} & \textbf{Time (s)} & \textbf{\#Non-Terminating} & \textbf{Time (s)}\\
         \hline
        \textsc{ULR-Baseline} & \multirow{3}{*}{3531} & 1309.64
 & \multirow{3}{*}{1088} & 47.13 \\
        \textsc{P-ULR-Seq} & & 492.61  & & 1.78  \\
        \textsc{P-ULR-Par4} & & \textbf{470.74} & & \textbf{1.62} \\
        \hline
    \end{tabularx}
    \caption{Comparison of \textsc{P-ULR} with \textsc{ULR} on 4619 lasso programs.
    For each tool, we report the number of instances proved terminating and their
    accumulated solving time, followed by the number of instances proved
    non-terminating and their accumulated solving time, both in seconds.}
    \label{tab:comparison}
\end{table}

\noindent\textbf{Experimental Setup.} The experiments were conducted on a machine with 16 Intel Core i5-14400T processors and 30~GB of memory. We used C program instances drawn from the \textsc{SV-COMP 2025}\footref{fn:svcomp25} and \textsc{TermCOMP 2025}\footnote{\url{https://termination-portal.org/wiki/TPDB}} competitions, as well as Boogie instances from the \textsc{Ultimate}\footref{fn:lassoranker} repository.

To obtain the benchmark, \textsc{Ultimate BuchiAutomizer} --- which relies on \textsc{ULR} as its internal lasso-level analysis library, also used by \textsc{Ultimate Automizer} --- was run on each input program with a time-out of 600~seconds, during which the lasso programs generated internally by \textsc{Ultimate BuchiAutomizer} and analyzed by \textsc{ULR} were serialized to JSON files. The time spent by \textsc{ULR} on each individual strategy was logged directly into the corresponding lasso file (\textit{excluding JVM startup time}). \textsc{P-ULR} was then run on the same benchmark with a time-out of 600~seconds per instance.
\\

\noindent\textbf{Sequential Comparison.} To ensure a fair comparison between \textsc{P-ULR} and \textsc{ULR}, we disabled the parallel execution of \textsc{P-ULR} and ran it in sequential mode, denoted \textsc{P-ULR-Seq}. On the other hand \textsc{ULR}, renamed as \textsc{ULR-Baseline}, executes its strategies sequentially in a fixed order: \textit{fixpoint} checking is launched first; if no conclusive result is obtained, \textit{GNTA} is run; and if still inconclusive, termination synthesis via ranking templates is performed in the following order: \textit{affine}, \textit{lexicographic}, \textit{multiphase}, \textit{nested}, \textit{piecewise}, \textit{parallel}, and \textit{composed lexicographic}. \textsc{P-ULR-Seq} follows the same strategy order. Both tools were evaluated using \textsc{Z3} as the underlying SMT solver.

Table~\ref{tab:comparison} reports the results of analyzing 4619 lasso programs, shown in the first two rows of the table. For each tool, we report the number of instances proved terminating and non-terminating, along with the accumulated solving time over the commonly solved instances. Overall, \textsc{P-ULR-Seq} achieves a lower solving time than \textsc{ULR-Baseline}, with a total gain of 862~s ($\sim$14~minutes). Notably, the accumulated time for 
non-termination synthesis is reduced by a factor of 26, from 47.13~seconds down 
to 1.78~seconds.

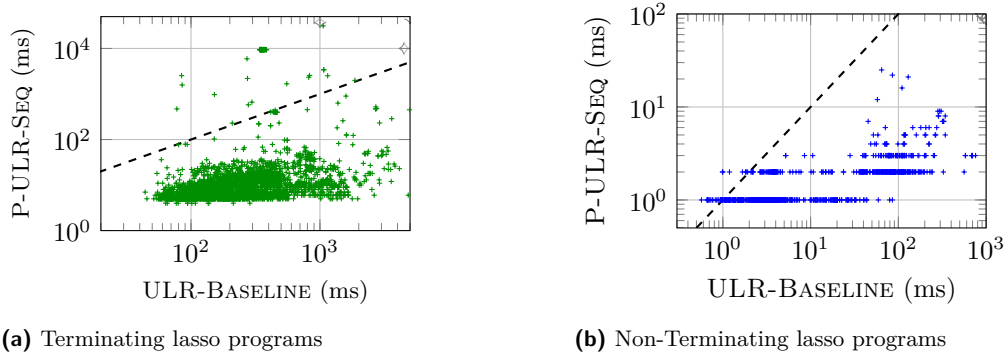
\begin{figure}[tp]
    \begin{subfigure}{0.45\textwidth}
    \begin{tikzpicture}
    \begin{axis}[
        width=0.9\textwidth,
        height=0.7\textwidth,
        xlabel=\textsc{ULR-Baseline} (ms),
        ylabel={\textsc{P-ULR-Seq} (ms)},
        grid=major,
        label style={font=\small},
        xmode=log,
        ymode=log,
        xmin=20, xmax=5000,
        ymin=1, ymax=50000,
        log basis x=10,
        log basis y=10,
        legend style={at={(1.3,1)}, font=\scriptsize},
    ]
    
    \addplot[
        scatter,
        only marks,
        mark size=1pt,
        scatter src=explicit symbolic,
        visualization depends on={\thisrow{nodes}\as\myvalue},
        scatter/classes={
            TERMINATING={mark=+,darkgreen}
        }
    ]
    table[x=Ultimate-all,y=pasttel,meta=label]
    {terminate_wst26_latex.dat};
    
    \addplot[
        domain=1:50000,
        dashed,
        thick
    ]{x};
    
    \end{axis}
    \end{tikzpicture}
        \caption{Terminating lasso programs\label{fig:term}}
        \label{fig:term-no-part}
    \end{subfigure}
    \hfill
    \begin{subfigure}{0.45\textwidth}
    \begin{tikzpicture}
    \begin{axis}[
        width=0.9\textwidth,
        height=0.7\textwidth,
        xlabel=\textsc{ULR-Baseline} (ms),
        ylabel={\textsc{P-ULR-Seq} (ms)},
        grid=major,
        xmode=log,
        ymode=log,
        xmin=0.3, xmax=1000,
        ymin=0.5, ymax=100,
        log basis x=10,
        log basis y=10,
        legend style={at={(1.3,1)}},
    ]
    
    \addplot[
        scatter,
        only marks,
        mark size=1pt,
        scatter src=explicit symbolic,
        visualization depends on={\thisrow{nodes}\as\myvalue},
        scatter/classes={
            NONTERMINATING={mark=+,blue}
        }
    ]
    table[x=Ultimate-all,y=pasttel,meta=label]
    {nonterminate_wst26_latex.dat};
    
    \addplot[
        domain=0.5:1000,
        dashed,
        thick
    ]{x};
    
    \end{axis}
    \end{tikzpicture}
    \caption{Non-Terminating lasso programs}
    \label{fig:non-term-no-part}
    \end{subfigure}
    \label{fig:without-partitioneer}
    \caption{Scatter plots of \textsc{P-ULR-Seq} against \textsc{ULR-Baseline}}
\end{figure}

The scatter plots in Figures~\ref{fig:term-no-part} and \ref{fig:non-term-no-part} illustrate the same results on a logarithmic time scale.

-- Regarding terminating instances (Figure~\ref{fig:term-no-part}), \textsc{P-ULR-Seq} significantly reduces the time required to find a ranking function compared to \textsc{ULR-Baseline}. The majority of lasso programs admit \textit{affine} ranking function as their termination proof, while a smaller subset requires \textit{nested}, \textit{lexicographic} or \textit{multiphase} ranking templates. Although \textsc{P-ULR-Seq} currently supports only \textit{affine} and \textit{nested} templates, this does not prevent it from producing termination arguments for the benchmark instances.

However, both tools do not always synthesize identical ranking functions. \textsc{ULR} applies an additional partitioning step that decomposes the problem into smaller sub-problems, enabling the synthesis of simpler and more readable ranking functions. \textsc{P-ULR}, which does not incorporate this step, produces ranking functions with larger coefficients and is currently restricted to \textit{affine} and \textit{nested} templates (a limitation discussed in Section~\ref{sec:implementation}). Furthermore, we observe that this partitioning step also reduces the execution time of \textsc{ULR-Baseline} on terminating instances, as reflected by the data points above the dashed $y = x$ diagonal.

-- Regarding non-terminating instances (Figure~\ref{fig:non-term-no-part}), two clusters are visible. The bottom-left cluster corresponds to instances solved via fixpoint-based non-termination proofs, where the solving times of both tools are closely aligned. The upper-right cluster corresponds to instances solved via geometric non-termination arguments (\textit{GNTA}), where \textsc{P-ULR-Seq} demonstrates a clear advantage in handling more complex non-termination proofs.
\\

\noindent\textbf{Parallel Comparison.} We now compare both tools in their full execution mode. \textsc{P-ULR} is run with 4 concurrent threads, referred to as \textsc{P-ULR-Par4}, each running an independent \texttt{SequentialAnalyzer} instantiated with one of the following strategies: \textit{fixpoint}, \textit{GNTA}, \textit{affine}, \textit{nested} (iterating from \textit{2-nested} to \textit{5-nested}) ranking templates. 

The results are reported in the third row of Table~\ref{tab:comparison}. \textsc{P-ULR-Par4} reduces the total execution time by a factor of $2.8$, achieving a gain of $884.4$~seconds over \textsc{ULR-Baseline}. However, compared to \textsc{P-ULR-Seq}, the parallel execution yields little additional speedup. This is explained by the composition of the benchmark: $98\%$ of lasso programs admit an \textit{affine} ranking function, which \textsc{P-ULR-Seq} already resolves first in its sequential order, leaving limited opportunity for parallelism to bring further improvement on this benchmark.

\section{Conclusion}
We presented \textsc{PaSTTeL}, a modular and tool-agnostic framework that unifies state-of-the-art termination and non-termination analyses for lasso programs into a concurrent portfolio. As a proof of concept, we instantiated the framework with the techniques underlying \textsc{Ultimate LassoRanker}, yielding competitive results. Future work includes addressing the limitations discussed above, as well as integrating recurrence set synthesis as an additional non-termination strategy to further enrich the portfolio.



\bibliography{biblio}

\appendix

\end{document}